# Elementary Interactions

## An Approach in Decision Tool Development

Heinrich Söbke [1 [0000-0002-0105-3126]] and Andrea Lück[1]

[1] Bauhaus-Institute for Infrastructure Solutions (b.is), Bauhaus-Universität Weimar, Weimar, Germany

`{heinrich.soebke|andrea.lueck}@uni-weimar.de`

**Abstract.** Multi-Criteria Decision Analysis (MCDA) is an established methodology to support decision making of multi-objective problems. For conducting a MCDA, in most cases a set of objectives (SOO) is required which consists of a hierarchical structure with objectives, criteria and indicators. The development of a SOO may require high organizational effort. This article introduces elementary interactions as a key paradigm for the development of a SOO. Elementary interactions are self-contained information requests that can be answered with little cognitive effort, which are made and processed with the help of a web platform. The pairwise comparison of elements in the well-known Analytical Hierarchical Process (AHP) is an example for such an elementary interaction. Each elementary interaction contributes to the stepwise development of a SOO. Based on the hypothesis that a SOO can be developed exclusively with elementary interactions, a platform concept is described. Essential components of the platform are a Model Aggregator, an Elementary Interaction Stream Generator, a Participant Manager and a Discussion Forum. The platform concept has been evaluated in a pilot study using a web-based prototype. The evaluation results demonstrate the general functionality of the platform concept. In summary, the proposed concept demonstrates the potential to advance the development of sets of objectives for MCDA applications: (1) The platform concept does not restrict the application domain, (2) it is intended to work with little administration efforts, (3) it lowers the organizational effort for developing a SOO. (3) it supports the further development of an existing SOO in the event of significant changes in external conditions. (4) The development process of the SOO can be recorded by the platform and thus becomes retraceable. The reproducibility may have a positive effect on the spread of MCDA applications. The traceability and the use of elementary interactions make the platform appear to be a suitable medium for Citizen Science-based approaches to the development of MCDA applications.

## 1 Introduction

MCDA is a group of decision support approaches which analyses multi-objective problems (Belton and Stewart 2002). In MCDA modelling aspects, such as stakeholder involvement and social participation are not essential for MCDA modelling, but are considered outcome-enhancing (Keeney 1996; Belton and Stewart 2002; Hendriksen et al. 2011; Marttunen et al. 2015). Thus, multiple MCDA variants integrate stakeholder engagement. Among these variants are the decision analysis interview approach (Marttunen et al. 2015), stakeholder multi-criteria decision aid (Banville et al. 1998), participatory Analytical Hierarchy Process (AHP) (Antunes et al. 2011), decision conferencing (Phillips 1989) and multi-actor multi-criteria analysis (MAMCA) (Macharis et al. 2009). In general, stakeholders can be involved in many stages of an MCDA development process (Lahdelma et al. 2000; Hendriksen et al. 2011; Lück and Nyga 2017).

These applications advantageously integrate software tools in their development processes. Examples of this would be the Decision Analysis Interview approach (Marttunen and Hämäläinen 1995; Karjalainen et al. 2013) and decision conferencing (Phillips and Bana E Costa 2007).

The development process of an MCDA application itself is demanding. A result of adhering to the requirement of integrating many stakeholder groups with diverse backgrounds into a joint, transdisciplinary process. Such an approach requires balancing various levels of cognitive skills, habits and cultures (Walter et al. 2008). For example, involved citizens and experts form a sharp contrast in terms of specific knowledge and experiences (Hendriksen et al. 2011). The modelling process is, also, prone to behavioral effects such as group interaction and influences by the facilitator based on the communication with the group (Hämäläinen et al. 2013). Moreover, MCDA development processes are commonly considered as very time- and effort-consuming (Lienert et al. 2013; Marttunen et al. 2015; Lück and Nyga 2017).

This article introduces the concept of a web-based software platform as a medium for the participatory development of an MCDA application involving all stakeholders. The central principle is the use of short interactions (between the participants and the platform. Participation from any location is enabled by the provision of the platform via the web. Time independence is enabled by the capability of asynchronous work, i.e., participants are not required to be online at the same time. Furthermore, time requirements for participation are flexible. Together these enable a large number of participants to contribute in the development of an MCDA application. Further, negative group effects should be avoided. The description of the platform is limited to the participatory creation of a set of objectives (SOO) as the core of a MCDA application.

This article is structured as follows: In the next section the theoretical foundations of the software-supported participatory development of SOO development is outlined. The concept of the envisioned platform is described in the succeeding section. Section 5 describes a pilot study based on the platform concept, whereas section 6 discusses the results. The article is concluded with a summary and the conclusions.

## 2    Theoretical Background

### 2.1    Participatory MCDA

Stakeholder involvement and participation is affirmed in MCDA literature (Saaty 1990; Keeney 1996; Banville et al. 1998; Lahdelma et al. 2000; Munda 2004); it allows incorporation of stakeholders' knowledge and values and enhances to bring structure to the planning, create discussion frameworks, and learning among stakeholders (Marttunen et al. 2015).

A variety of participatory methods is discussed (Petkov et al. 2007; Salo and Raimo 2010; Gabriel et al. 2016; Lück and Nyga 2017). These methods range from workshops, stakeholder groups meetings, interviews, written surveys, brainstorming and writing, morphologic analysis, literature research, and panel of experts (Palme et al. 2005; Hendriksen et al. 2012; Domènech et al. 2013; Marques et al. 2015; Marttunen et al. 2015; Lück and Nyga 2017). The application of such methods is time and staff resource consuming (Lienert et al. 2014; De Brito and Evers 2016; Lück and Nyga 2017). By applying participatory approaches in presence meetings (e.g., workshops, sessions, panels), there may occur strategic, tactical, social, and psychological issues in the decision modelling process faced by individuals (Kilgour et al. 2010). Negative effects like dominance of stakeholders (Hsu and Sandford 2007), strategic answers by stakeholders (Jonsson et al. 2007) and groupthink phenomenon (Kerr and Tindale 2004) have been observed.

There are structured communication techniques, such as the Delphi-method (Hsu and Sandford 2007), which aim to reduce negative group effects by employing repeated questionnaires and aggregating facilitators for achieving consensus. The proposed approach aims at reducing negative group effects. However, it accentuates asynchronous activities, algorithm-based aggregation of answer and the inclusion of all stakeholder-groups while not requiring personal meeting.

### 2.2    Set of objectives (SOO) development

For conducting a MCDA, the following steps are usually accomplished: (1) clarify the decision context; (2) define objectives and attributes; (3) develop alternatives; (4) estimate consequences; (5) evaluate trade-offs and select alternative, and (6) implement, monitor and review (Gregory et al. 2012).

The development of a SOO is carried out in the first two stages according to Gregory et al. (Gregory et al. 2012) and includes the definition of the assessment goal and the collection of objectives and criteria[1] (Lück and Nyga 2017; Nyga et al. 2018). The assessment goal is divided into objectives. Each objective is specified in more detail by so-called criteria. A criterion is measured by indicators, which provide concrete values. Fig. 1 depicts the general structure of a SOO.

Besides weights, the SOO is supplemented by transfer functions to serve as the basis for an MCDA application. The SOO represents objective aspects of the MCDA application, while the weights represent the subjective preferences. An example of this MCDA application differentiation of objective facts and subjective preferences is the nuclear accident of Fukushima. No facts were changed by the accident, but the preferences of the citizens have evolved and led to an exit from nuclear energy in Germany (Hermwille 2016; Renn and Marshall 2016).

---

[1] The terms "attribute" and "criterion" are used synonymously.

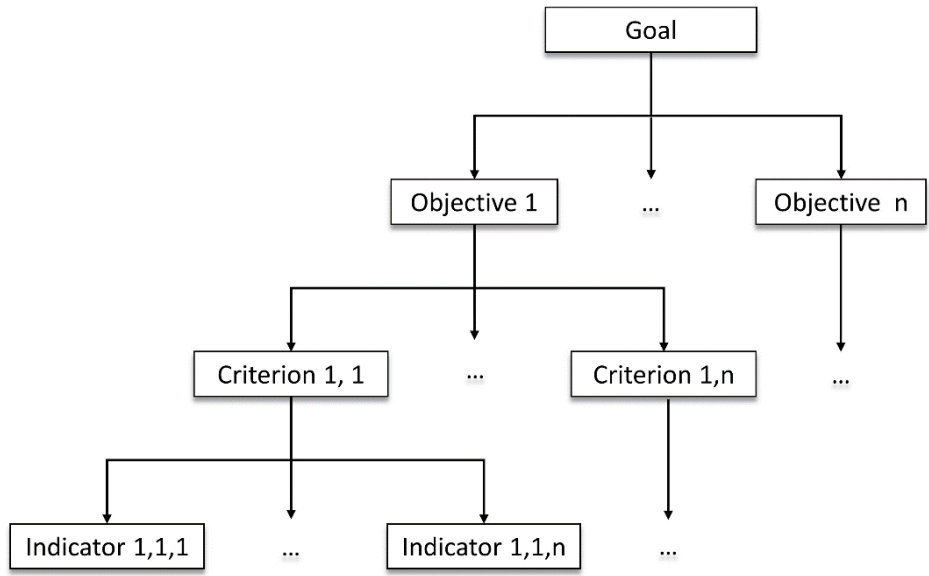

**Fig. 1.** Generalized Structure of Set of Objectives

### 2.3    Participatory MCDA with software tools

There are many applications of MCDA software (Buede 1992; Buede 1996; Vassilev et al. 2005; Ishizaka and Nemery 2013; Oleson 2016; Weistroffer and Li 2016; Mustajoki and Marttunen 2017) as well as many case studies (Korosuo et al. 2011).

Marttunen et al. (Marttunen et al. 2015) discusses a list of potential problems occurring during personal interactive interviews with MCDA software. He argues that the software-based MCDA modelling requires time and commitment from stakeholders, problems of understanding or accepting the method and its principles by some participants, support by experienced decision analyst is required and the potential for unintentionally influence of interviewees answers may occur (Marttunen et al. 2015).

The proposed approach is preventing these problems as the user can choose their engagement level on their own, the elementary interactions (EIs) do not require a deeper understanding of MCDA modelling and the EI can be adopted to user's abilities. Furthermore, there is no decision analyst who's influencing the process.

Mustajoki and Marttunen (2017) provide a recent survey of MCDA software, especially in the context of environmental planning processes Mustajoik and Marttunen state that "[there] are numerous MCDA software tools available". Most of the software packages investigated support MCDA-related models and the elicitation of preferences via questionnaires. However, the development of a SOO with the help of a large number of participants is not mentioned. "We think that none of the software tools in our analysis is such that users without any prior experience of MCDA could use it." The proposed platform concept ensures that the initiators of an MCDA application only are required to be trained in the usage of the platform, while the participants simply have to perform self-explanatory elementary interactions with the platform.

## 3   The Concept of Elementary Interactions

Elementary Interactions (EIs) are the central construct of the platform concept. EIs are defined as short participant-platform interactions. Ideally, EIs are closed questions in which the participant must choose from a predefined set of answers. EIs are self-contained and require short human processing time only, i.e., they are accomplishable with a few clicks or typing a term in less than a minute. Thus, the platform creates a low-threshold for participation in the development process.

Figure 2 shows three examples of website components asking for short interactions, the inspiration for the elementary interactions proposed here. The requested interactions require the participant to make a short decision and externalize this decision with one click. Although it is not possible to restrict the EI's cognitive complexity to such a low level (e.g., confer EI Name in **Fehler! Verweisquelle konnte nicht gefunden werden.**, which requires identification of a meaningful word and typing it in), it is considered as an essential design trait for EI. The goal is to keep the level of cognitive complexity as simple as possible for enabling working on EI on the fly and encourage complex operations. A method for limiting the level of cognitive complexity is the utilization of closed questions. An example is the question "Is criteria A or criteria B more important in order to measure objective C?" (This kind of question is well-known as paired comparison from the priority evaluation within the AHP (Saaty 1990)). Using such a design allows short feedback cycles: a participant is given a short task that can be completed in a second and for which feedback is immediate. This should tempt the participant to the next elementary interaction, which is just as easy to accomplish. This principle of a stream of elementary interactions can be observed, for example, in surveys conducted in the field of public opinion research by the company Civey (Wurnig 2017; Civey GmbH 2018). Participants can stop answering elementary interactions at any time.

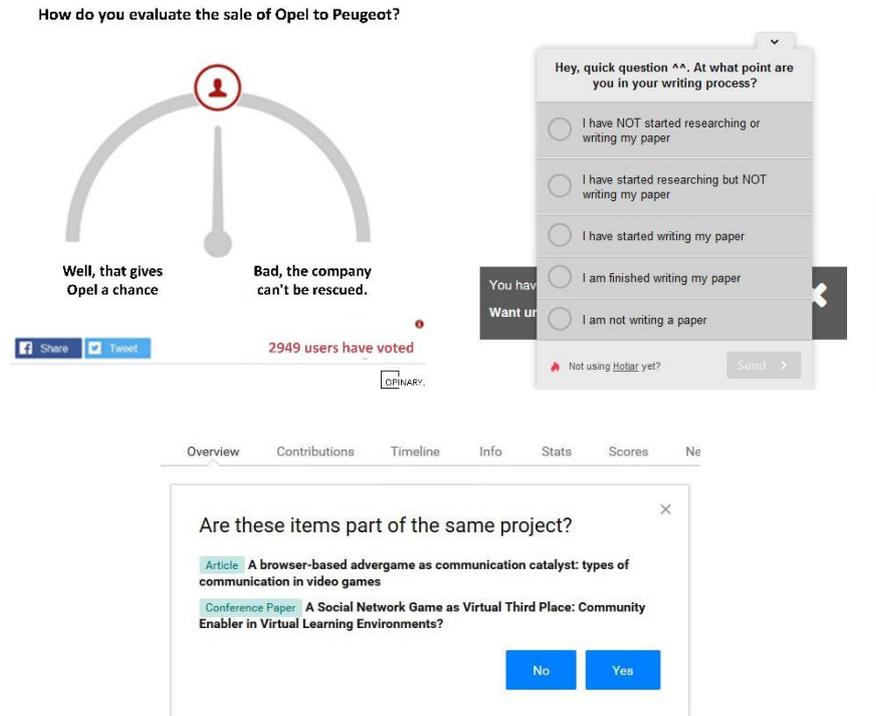

**Fig. 2.** EI examples: *Top left*: request for seamless personal evaluation of a company takeover (Opinary GmbH 2017; SPIEGEL ONLINE GmbH 2017); *top right*: slide-in single choice question for the reason of web page visit (EasyBib 2017); *bottom*: in-passing request for additional attributes of content in a domain specific content management system (ResearchGate GmbH 2017)

## 3.1 Elementary Interaction Types

EI must fulfill different purposes such as for creating SOO elements, structuring, or validation. In the following, the EIs are categorized by purpose and described with the help of examples. The list of purposes represents a draft and is not complete. EI are summarized in **Fehler! Verweisquelle konnte nicht gefunden werden.**.

**Table 1.** EI Schema Description

| Schema Element | Description |
|---|---|
| Id | Id of the EI |
| Description | Describes the context and purpose of the EI |

| Category | The category refers to the purpose an EI serves. Commonly, there can be different EI for achieving one purpose, e.g. there is more than one EI to validate an element. |
|---:|:---|
| Elements Affected | Names the elements of the assessment model to which this EI is applicable (e.g. Objective, Criterion, Indicator) |
| Impact | The Impact is described here. |
| Sample Question | A sample question that illustrates the EI. |
| Interaction | The action the user has to take to fulfill the EI |

**EI Category *Create*.** The first necessity is to ask the participants for appropriate SOO elements. This is accomplished by the EI *Name* (cf. **Fehler! Verweisquelle konnte nicht gefunden werden.**, Id 1), which asks for example: *"Please name a criterion, which is important to assess the objective time."* After having been answered by multiples users, EI Name results in a set of potential elements (*Element Candidates*) This EI is considered as cognitively complex, because the participants have to think creatively about a suitable term, which, for example, designates a criterion, and must also type in the term.

**EI Category *Validate*.** As soon as an element has been named, it has to be validated. This is the goal of another EI *Confirm* (cf. **Fehler! Verweisquelle konnte nicht gefunden werden.**, Id 2): The participant is asked if a given element candidate must be considered as an element, e.g., "Is *direct costs* a valid criterion to assess the objective *economy*?" If an element candidate reaches a certain validity level, generation of elements of the subordinated level can be started, e.g., if a criterion has been validated, suitable indicators can be generated. The validation of SOO Elements requires support by appropriate validity measures. For example, the percentage of confirmations compared to the rejections of a SOO element.

**EI Category *Structure*.** The goal of structuring criteria and objectives is the identification of duplicates and an hierarchical structure. The EI *Identify duplicates* (cf. **Fehler! Verweisquelle konnte nicht gefunden werden.**, Id 5) works on two random elements. It helps to discover duplicates and elements with similar meanings. If the results of this EI point to two (or more) potentially similar elements, the EI *Determine Common Name* requires the participant to enter a common name. If a provided name achieves a defined validity (resulting from confirming EIs similar to EI *Confirm*), the underlying similar elements are removed from the model and the resulting model is added. Further EIs evaluate the need to restructure the elements hierarchy. The EI *Select parent element* (cf. **Fehler! Verweisquelle konnte nicht gefunden werden.**, Id 8) challenges the current assignment of an element (criterion or indicator) to its parent, e.g., "*What is the most appropriate objective for the criterion 'direct costs': 'economical objectives', 'environmental objectives' or 'social objectives'?*". The answers to this EI either confirm the assignment, provide hints to relocate it or identify new elements of the superordinate level.

**EI Category *Determine Weights*.** The determination of weights is giving a priority to the elements of SOO. An example is the pairwise comparison, accomplished by using EI *Prioritize pairwise* (cf. **Fehler! Verweisquelle konnte nicht gefunden werden.**, Id 3), e.g. *"Is the objective '*direct costs*' more important than '*indirect costs*?"* (measured on a Likert-scale). A variant of this EI is the specification of more than two answer options. The EI *Choose set-based* (cf. **Fehler! Verweisquelle konnte nicht gefunden werden.**, Id 4) implements multiple answer options: *"Which five of the following criteria are the most important criteria for measuring economic objectives of a water infrastructure system?"*.

**Table 2.** Overview of Elementary Interactions

| ID | Name | Description | Cat. | Elem. | Purpose | Sample question | Interaction |
|----|------|-------------|------|-------|---------|-----------------|-------------|
| 1 | Name | Used to add new elements to the model. Therefore, it requires the explicit naming of such an element. | Element creation | Objective, Criterion, Indicator | Adds a new element of the given type to the model. | Please name a criterion, which is important to assess the objective time. | Typing in a name |
| 2 | Confirm | This EI is used to validate an element of a model by asking a user for confirmation. | Validate | Objective, Criterion, Indicator | Increases the validity. | Is *Direct Costs* a valid criterion to assess the objective *Economy*? | Choosing confirmation or rejection |
| 3 | Prioritize pairwise | This EI is used to prioritize an element of a model over another by asking a user. | Validate, Weigh | Objective, Criterion, Indicator | Weighs, Increases the validity. | Which criterion is more important to describe the objective *Economic Objective*? *Direct Costs* or *Indirect Costs*? | Choosing one of two choices. |
| 4 | Choose set-based | This EI is used to select the most relevant elements of a set. Depending on a customization, the selection may be ordered or unordered. It preferably should be implemented via Drag and Drop in a graphical user interface (GUI). | Validate, Weigh | Objective, Criterion, Indicator | Increases the validity. | Which five of the following criteria are the most important Criteria for measuring Economic Objectives of a Travel Type? [Select in order of importance] | Choosing up to five of the given set of elements. |

| 5 | Identify duplicates | This EI identifies duplicate elements, which may differ in names, but probably have the same meaning. | Validate, Restructure | Objective, Criterion, Indicator | Increases the validity. | Do you think, "Indirect Costs" and "Direct Costs" are the same criterion? [To which extent do the criteria "Indirect Costs" and "Direct Costs overlap?] | Answering with *Yes* or *No*. A variant of this EI could ask for the grade of identity on a scale from 0 to 100%. |
|---|---|---|---|---|---|---|---|
| 6 | Determine common name | This EI asks the user for a common name for two or more elements. | Name | Objective, Criterion, Indicator | Determines the validity of an element resp. restructures the model, if a threshold validity has been reached. | What is a common name for the criteria "Direct Costs" and "Indirect Costs"? | Entering a name. |
| 7 | Select parent element | This EI asks the user for the appropriate parent element. It offers all available parents of the hierarchy level of its parent and lets the user choose the most appropriate parent element. | Validate, Restructure | Criterion, Indicator | Determines the validity of an element resp. restructures the model, if a threshold validity has been reached. | What is the most appropriate objective for the criterion "Direct Costs": Economical Objectives, Environmental Objectives or Social Objectives? [Provide an alternative objective, if no suggestion fits really well.] | Choosing one of multiple choices [or entering the name of an alternative]. |

## 4    Platform concept

The proposed platform consists of seven elements (see Fig. 3). The simulation model (2) represents a system of the real world (1). Based on the interactions of participants (3) with the platform, the set of objectives designer (4) creates the SOO and weights (5) with the help of so-called elementary interactions. In the end, the assessment result (6) serves as a basis for decisions (7).

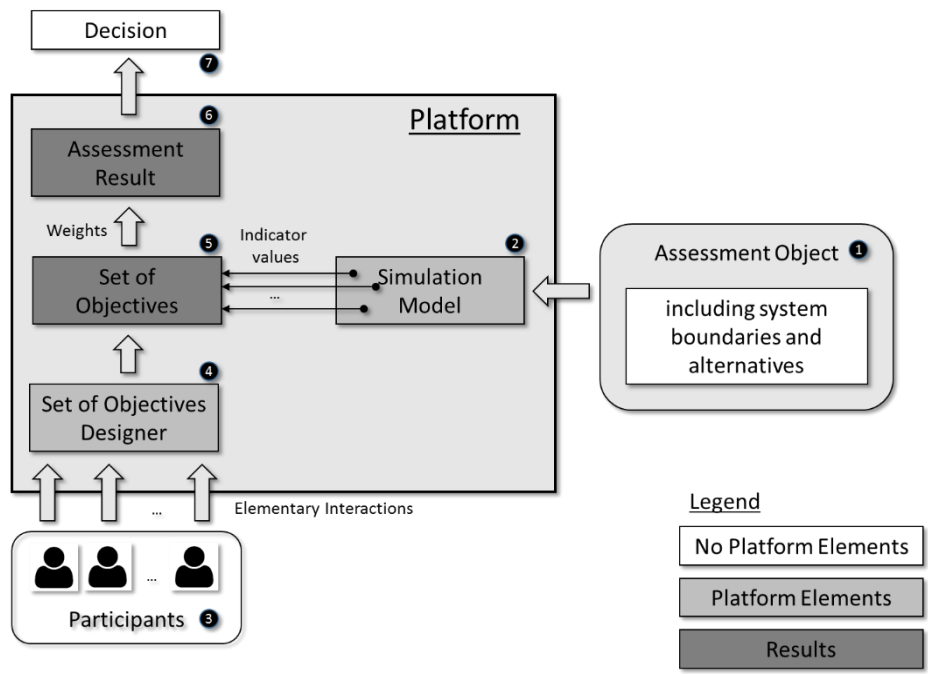

**Fig. 3.** System Perspective

(1) The *assessment object* includes the system boundaries and the alternatives for the assessment objective.

(2) Each indicator of the SOO must be calculated and either requires an algorithm or manual data input, e.g., in the case of an expert estimation. The input values for calculation of the indicators are stored in the *simulation model*, which represents a model of the real-world system to be evaluated. The development of an indicator is always accompanied by the modelling of suitable attributes in the simulation model. Thus, when an algorithm is defined – by means of an expression editor, which can explore the underlying model – it relies on the already present attributes or it adds new attributes to the simulation model. The simulation model grows in parallel with advancing SOO development. This means that both the meta-model of the simulation model is developed as well as corresponding values for concrete assessment object examples are provided. At this point of the process, a lack of data may emerge and may require a redesign of indicators and their algorithms.

(3) *Participants* are required for the functioning of the platform. They are managed through the participant manager, but they are not seen as part of the proposed platform as such.

(4) The *Set of Objective Designer* is collecting the information given by the participants through EIs. This includes the collection and structuring of objectives, criteria and indicators. Furthermore, weighting of SOO is conducted by *Set of Objective Designer.*

(5) The *set of objectives* results from *set of objectives designer* and *the simulation model.* Both components, their interactions and the development of SOO are explained in sections 4.2 and 4.3.

(6) The combination of SOO Milestone with corresponding weights and the example data of the simulation model allows a suggestion for an *assessment result.*

(7) Based on the assessment results, a *decision* can be made.

## 4.1 Use Case

To illustrate the intended workflow of the platform, the following use case includes the relevant steps using the example of creating a SOO for a water infrastructure MCDA application.

**Step 1: Definition of the assessment goal, selecting and activating platform participants.** One or more persons – the initiators - recognize the need for an MCDA application. They define the goal and the system boundaries of the real-world system. Further, the initiators identify relevant stakeholder groups. In the case of water infrastructure, typical stakeholder groups have been identified before (Lienert et al. 2013; Lück and Nyga 2017). To reach a large number of potential participants, related associations to the assessment topic should be identified.

**Step 2: Starting the development process.** As soon as an invited participant creates an account on the platform, he is able to inform about the purpose and aims of the MCDA application. During this introduction, the participant answers multiple choice questions. These questions inform the participant about the context and asses the status of the participant's knowledge. Thereafter, the participant can browse through the current SOO (which at the beginning comprises the goal only). Alternatively, the participant can answer a sequence of elementary interactions. The sequence is created on a semi-random base. The participant can stop answering to EIs at any time. Dependent on the status of SOO, not all proposed EI may be available yet. For example, if there are only objectives, requests for indicators are not yet possible, because indicators refer to a criterion.

**Step 3: Development process.** The development process for the SOO should run without the need for administrator intervention in most cases. Tasks such as evaluating the validity of the proposed elements of the SOO and generating the EI stream are performed with the help of algorithms. However, initiators must monitor activities on the platform and intervene in situations when there is a lack of participants or if the goal has not been defined clearly.

**Step 4: Evaluation of the resulting SOO.** After threshold values of validity have been reached, a milestone version of the SOO is created. This version of the SOO can be integrated into an MCDA application.

**Step 5: Evolution.** When external conditions have changed significantly (e.g. civic preferences), the developed SOO may not applicable any longer. In this case, the platform can be used for a further development of the SOO based on the already identified SOO elements of the platform.

In the following, specific aspects of the platform concept, which facilitates the implementation of the given use case, are highlighted. Among them are besides the information perspective, the concept of *elementary interactions,* the platform components *Set of Objectives Designer*, *Participant Manager* and *Model Aggregator*.

## 4.2 Set of Objectives Designer

The *set of Objectives Designer* is responsible for the development of a viable SOO and the assessment of weights of SOO's elements. Fig. 4 depicts the structure and workflow of the SOO-Designer. Central component is the *EI Stream Generator*. It creates elementary interactions based on multiple sources of information. First, the current SOO is analyzed for missing information. For example, if a criterion misses indicators, elementary interactions to survey indicators for the criteria are generated.

A further information source is the *Participant Manager*, who possesses information on participant competency. For example, if the *Participant Manager* has recorded little technical competency for a participant, it does not make sense to provide this participant with elementary interactions for naming subject-specific criteria. Rather, more elementary interactions should be asked about preferences for weighting of the criteria.

Participants' answers to the elementary interactions are delivered to the *Model Aggregator*, which integrates answers into the SOO. The *Model Aggregator* uses information provided by the *Participant Manager*. Besides competencies, this information comprises measures of reliability, which are used to provide weights to the answers. Further, the answers are used to update the participant's specific information of the *Participant Manager*; described in detail in the following section. The *Model Aggregator* is described in more detail in the next but one section.

An additional component of the *Set of Objectives Designer* is the *Discussion Forum*. A discussion forum of this kind can be realized with the help of software packages like MediaWiki (Wikimedia Foundation Inc. 2017) or Stack Overflow (Stackoverflow.com 2012). Whereby, discussions about the elements of the SOO between participants should be fostered to enable collaborative development. This component is integrated into the *Set of Objectives Designer* using hyperlinks: whenever an element appears in the graphical user interface, e.g., in the question of an EI, a hyperlink leads to the according description and discussion page of this element.

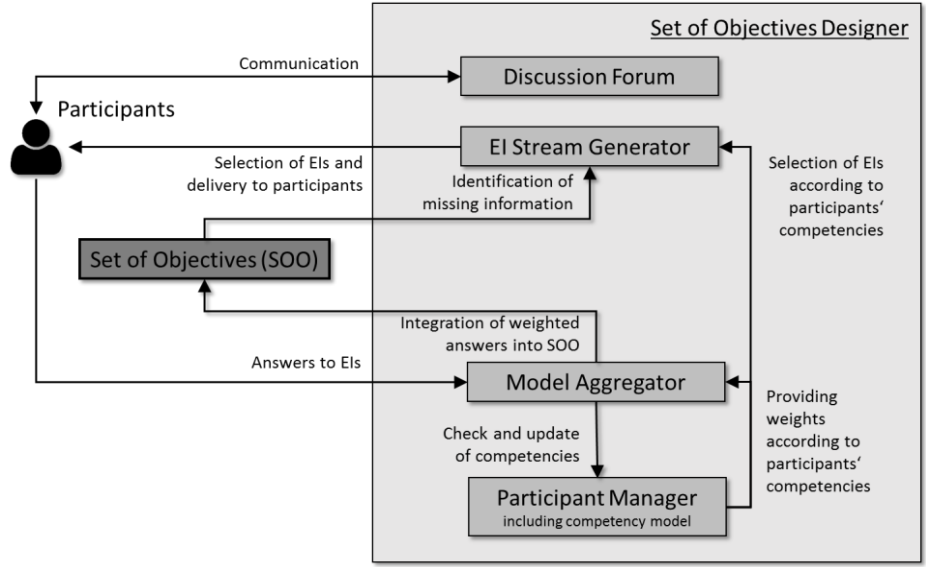

**Fig. 4.** Set of Objectives Designer: Components and Process

### 4.3 Participant Manager

In general, various stakeholder groups influence the design process of a MCDA system (Banville et al. 1998; Lienert et al. 2011; Lienert et al. 2013; Ferretti 2016). A possible enumeration of stakeholder groups includes decision makers, interest groups, experts and planers (Lück and Nyga 2017).

In the case of the proposed platform, initiators are a distinct group in the MCDA application design process. The initiators define the goal of the MCDA application, the system boundaries and invite potentially participators. They ensure all involved stakeholders are represented, i.e. that the entirety of platform users can provide specialist knowledge and preferences of affected stakeholders at the same time. Furthermore, end-users are a specific group that is subsumed under the term interest group in the above enumeration. End-users can be defined as stakeholders without specialist knowledge about the assessment object, who are impacted by an MCDA application-based decision. End-users in the context of water infrastructure are citizens.

It is necessary to estimate the role and the capabilities of each platform user. For example, the contributions of a proven expert to the SOO elements have to be more weighted than "guesses" of the end-users. Hence, a *competency model* is created and maintained during the platform operation. This model is used to weigh the impact of executed EIs. For example, the more expertise a participant demonstrates, the more impact will get the participant's contributions to SOO elements. Among possible influencing factors of such a user model are:

- **Assessment Results**: When a participant registers on the platform, an introductory test is done, which assesses the technical expertise of the user. The provision of an initial test has to be done by the initiators. During the design process, it can be extended, e.g., by a collaborative question design tool (McClean 2015).

- **Self-estimation**: If a participant identifies his- or herself as an end-user, the initial focus of the EIs may be set on contributing to preferences.
- **Reputation:** In crowdsourcing systems, contributors often are assigned an attribute *Reputation* (Adler et al. 2011), which is a measurement of the quality of their previous contributions to the system. At the same time, reputation is used to derive system permissions. An example is the Question & Answer software *Stack Overflow* (Stackoverflow.com 2012).

### 4.4 Model Aggregator

A characteristic of the platform is the continuing development process while the MCDA application is already capable of supplying an assessment result. This leads to the question, at which point of development such a model can be considered as stable. It is suggested to introduce various attributes, which each describe a validity measurement of an element. First, an attribute *validity* accumulates the element's validity. It determines for example if the name of the element is reasonable. Further, an attribute *validityStructure* holds a measure of the correct structural position of the element, i.e., if the element is located correctly in the SOO structure. Another measurement of validity can be the attribute *validityChildren* is a measure for the stability of the subordinated elements, e.g., if those elements define a complete set and are mutual independent.

The values of validity-describing attributes are continuously updated by EIs, which affect the related elements. For example, if multiple users name the same criteria via EI *Name* (e.g., *direct costs*), the validity of the element (represented by the attribute *validity*) is increased naming by naming. The confirmation of an element (EI *Confirm*) increases the value of this attribute, whereas a rejection decreases it.

In general, the question of validity occurs on at least three levels. The first level are elements: Validity of an element is indicated by the attribute *validity*. The next level are groups, which are the subordinated elements of a parent element, e.g., the criteria that belong to an objective. Group-based validity increases if elementary interactions of the category *Structure* do not result in changes: Each negation of an EI "Find Duplicates" increases the validity, each confirmation decreases it again. The children's attributes *validityStructure* contribute to the *validityChildren* attribute. The third level are tiers: There are two tiers here: the first one is the hierarchy of objectives, criteria and indicators, the second one is given by the weights of objectives and criteria. The determination of the weights is reasonable only, when the underlying first tier has been captured in a milestone, i.e. is no longer subject to changes. The decision, when the design of the first tear is completed, can be made by the platform automatically dependent on the tier's *validity* attributes: when the average validity of the elements has reached a threshold, a milestone is created. Thereafter, the step of determining weights is based on this milestone version of the elements hierarchy.

Formulas to calculate the values of validity attributes, have to take into account certain stipulations: An element's validity can be considered as stable, when its value does not change significantly during the last affecting EIs.

## 5    Case Study

The platform concept has been validated in a pilot study (Körting 2018). In the following, the study is described, results are presented, and discussed.

### 5.1    Study design

Based on paper prototypes, a platform-based study was performed with the goal to assess the *sustainability of water infrastructure*. The standard test functionality of the learning platform *moodle* (Moodle.org 2018) was used as technical basis. Moodle's test functionality allows to deliver questions randomly from various pools of questions. Participants have been recruited from the scientific staff of a chair of urban wastewater management (n=12) and from the acquaintances of the study lead (n=14). Altogether 26 participants were involved in the case study. At the end of the case study, 18 participants were still active. While the platform concept employs the concept of a continuous EI stream and model integration, the study followed for practical reasons a turn-based approach. In each turn, a set of EI-based question was generated (typically 10-20) manually by the study lead. Thereafter, participants were requested to answer a questionnaire, which consisted of 8-10 random EI-based questions. The answers to the EI-based questions were manually integrated into the SOO. The manual integration included the correction of spelling errors. The updated SOO was then used as the baseline for generating a set of EI-based questions for the next round. The main goals for the generation were to close gaps in the SOO as research objectives of the study  in detail, such as checking for duplicates or if an objective can be derived from existing criteria.
In total, 2,200 questions were answered, and the results were evaluated by the participants with feedback questionnaires. Altogether, 12 rounds were performed with the frequency of two rounds a week.

### 5.2    Results

The SOO achieved after 12 rounds is depicted in **Fehler! Verweisquelle konnte nicht gefunden werden.**. At a first glance, it can be concluded that the SOO is not yet complete. This applies to both the objectives and the indicators. A bottom-up approach was used to derive the objectives from the criteria. The objective *ecological objectives* was finally set so that criteria could be assigned. Further gaps (e.g., no indicators for the criteria in the right) in the SOO are due pursuit of specific research aspects.
A relevant study result is the specific order of employment of EI types. Starting from the goal, the first turn consisted of questions for naming criteria only. In the next turn, the criteria of the first round had to be validated and to be checked for duplicates. After the first criteria had been validated, first indicator-generating EIs could be issued. In consequence, the description schema of EIs requires the naming of prerequisites for application of the EI. Further, state models and state transition diagrams for SOO elements would be beneficial for the concept description.

The study allowed first experiences with validity measures. Heuristically, criteria validation requires a confirmation rate of 75 % or more and at least 10 confirmations. This validity measure was applied for criteria and indicators and helped to identify the SOO depictured in **Fehler! Verweisquelle konnte nicht gefunden werden.**.

Difficulties occurred when the names for criteria were not clear. As a result, many participants were overwhelmed by the EI *Confirm* and were unable to answer the question. Two reactions have been taken. Firstly, the option "I don't know" was added. Secondly, definitions for the criterion were requested. With the help of the EI *Choose set-based*, the participants were then able to determine a suitable definition for a criterion.

All participants received the same questions without any differentiation according to the stakeholder group. Due to informal feedback, technical questions were too demanding for some participants, especially those who identified themselves as interested layman. Therefore, the participants should be assigned to stakeholder groups and the questions should be stakeholder group specific. Further feedback of the participants contained statements about the repetition of questions.

Various question designs were used in the study. For the EI *Identify duplicates*, for example, in addition to the original Yes/No variant, the question of how far both elements overlap on a 7-point Likert scale was also raised. No clear results could be found, it seems that participant-type dependent preferences exist.

The study confirmed the basic functionality of the procedure and revealed tasks to be worked on. These tasks include adherence of the EI stream generator to the participants stakeholder group, the motivation of the participants, the required large sample size of participants, the question design, sound validity measures and the challenge to identify objectives.

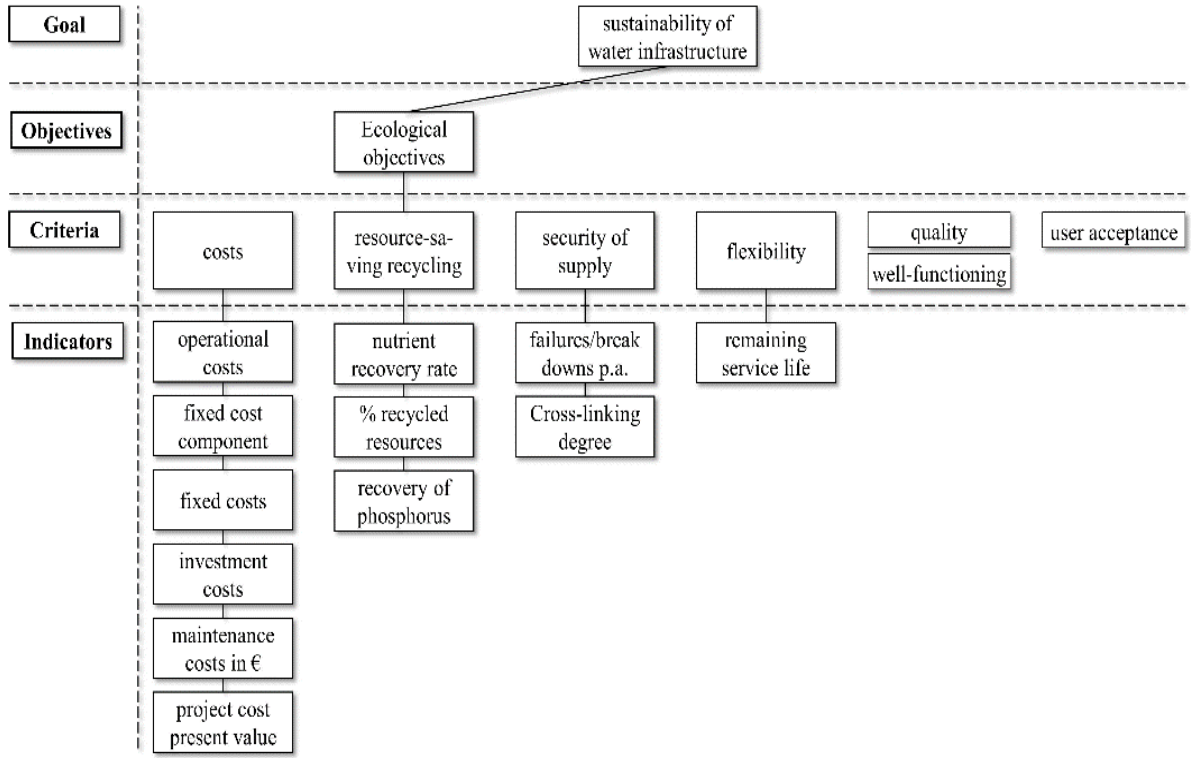

**Fig. 5.** Study Result: Preliminary Set of Objectives

## 6    Discussion

The described concept of a platform-mediated almost un-administered approach to create SOOs seems to be attainable. Such a platform enables the development of SOO and suitability of MCDA applications for various purposes. Further, it could be made openly available. The platform concept can modularly integrate already established methods of decision method development, such as stakeholder analysis, determination of weights and transfer functions.

An essential question to be answered is that of the possible participants in the development of a SOO using the platform. Citizen Science seems to be a promising application context. Although the term Citizen Science is not clearly defined (Kullenberg and Kasperowski 2016; Eitzel et al. 2017), two identifying marks are repeatedly referred to as characteristic of Citizen Science. First, Citizen Science is open to a great number of potential participants and, second, "intermediate inputs such as data or problem solving algorithms are made openly available" (Franzoni and Sauermann 2014). The proposed platform concept seems to fit well into the context of Citizen Science. It requires a great number of participants. Their recruitment is a common task in Citizen Science projects. In addition to breaking down the workload into small tasks - as the presented platform concept does - the temporal independence is an important characteristic of tasks that can be assigned to the participants of Citizen Science projects. (Sauermann and Franzoni 2015). In current Citizen Science projects, participants predominantly are occupied with simple tasks like collection of data or documenting observations (Wiggins and Crowston 2012; Sauermann and Franzoni 2015). Nevertheless, more complex tasks are observable in a few Citizen Science projects (Dolejšová and Kera 2017). As became apparent in the pilot study, the development of a SOO sometimes places high demands on the participants, which must be considered in the design of a Citizen Science process. A common challenge of Citizen Science-based projects is the phenomena of vandalism and content manipulation. This can be considered as a problem; however, countermeasures have been subject to research (e.g. Adler et al. 2011). The main motivation of organizing the development of MCDA applications as Citizen Science processes is not the facilitation of unused work capacities to reduce costs of scientific outcomes, as it has been observed (Sauermann and Franzoni 2015). Rather, employing Citizen Science in this context can be considered as a means of reducing the complexity of the development process to an operable level.

The platform concept relies on a great number of participants, as the pilot study has revealed. For Citizen Science projects, however, it is known that user activity decreases over time (Sauermann and Franzoni 2015). For this reason, it is necessary to motivate the participants over and over again. The proposed platform concept therefore needs to include methods of motivation design for the participants. Gamification, the application of gaming principles to real-world tasks (Deterding et al. 2011) is a methodology to foster motivation and engagement of the participants. The platform is expected to offer multiple opportunities of gamification: the platform generates a huge amount of usage data, e.g., the amount of interactions of each user or the number of consecutive days of logins. Especially, the introduction of a reputation system is considered as engagement fostering without affecting the participants results negatively (Thiel 2016; Thiel and Fröhlich 2018). Move over, immediate feedback is considered as an important means of fostering engagement (Garris et al. 2002). Immediate feedback can be given by an extensive statistics component, which would visualize the effects of any performed EI. Key figures, such as "Participant's elementary interactions" and "Platform elementary interactions in the last 24 hours" are motivating for a part of the participants.

In general, the platform concept enables the use of visualizations, since the available information is integrated on the platform. Visualizations are known as beneficial for cognitive processing of information, especially when combined with interactions (Liu and Stasko 2010). In particular in multimedia learning, visualizations are attributed a prominent role (Mayer 2009). To visualize the results of MCDA applications there are already various approaches, especially to compare different variants of diagrams (Lami et al. 2014; Miettinen 2014; Haara et al. 2018). These capacities can be further supplemented within this platform by the integration of all important information over time, as well as the possibility for interactive visual evaluation of the platform-contained information, such as performing a sensitivity analysis. Further, the capacity of the platform to trace the changes of various components over time, such as the simulation model of the real-world systems and the preferences model, would support the visualization of these changes over time.

The platform concept relies on elementary interactions of low cognitive complexity. As the pilot study has shown, it is not possible to limit the cognitive complexity of elementary interactions consistently to the level of simple multiple-choice questions. For example, a multiple-choice question regarding the best definition of a SOO element requires a lot of reading work. Another example is the creative work required when naming new elements. Therefore, further research must clarify what degree of cognitive complexity is operable for elementary interactions without perceiving them as hard work so that participants are not discouraged from working on them.

## 7  Conclusion

The development of MCDA applications is due to consistent findings in literature a complex process, which requires high organizational efforts. This article describes a platform concept for developing sets of objectives (SOOs). A SOO is an essential component of an MCDA application. Key paradigm of the concept is the decomposition of design decisions into short interactions, so-called *elementary interactions* (EIs). Based on the information collected by these elementary interactions, a structured SOO consisting of objectives, criteria and indicators evolves over time. Relevant components of the platform concept are a *Participant Manager*, which holds a competency model for each participant, a *Model Aggregator*, which transforms the answers received by elementary interactions into the SOO, an *Elementary Interaction Stream Generator*, which creates streams of elementary interactions due to the information required for completing the SOO and suitable for each participant, and a *Discussion Forum* to foster communication between participants.

A pilot study confirmed the general functional capability of the platform concept. However, it also helped to identify further research demands, such as determining methodologies to cluster criteria into objectives and exploring the cognitive complexity of elementary interactions.

In summary, the platform concept offers the following advantages: (1) The platform concept is open to any MCDA application domain, (2) it is intended to work with little administrative effort, (3) it lowers the organizational effort for developing an SOO, (4) it supports the further development of an existing SOO in the event of significant changes in external conditions, and (5) the development process of the SOO can be recorded by the platform and thus becomes retraceable. The reproducibility may have a positive effect on the acceptance of MCDA applications. Combined, traceability and use of elementary interactions make the platform appear to be a suitable medium for Citizen Science-based approaches to the development of MCDA applications. Hence, the platform concept has the potential to significantly expand the variety of methods for creating MCDA applications.